\documentclass[12pt]{article}
\usepackage{latexsym, graphicx,cite}
\usepackage{amsmath}
\usepackage{amssymb}
\usepackage{amsthm}
\allowdisplaybreaks

\usepackage[top = 1. in,bottom = 1. in, left = 1 in, right=1 in]{geometry}

 \newcommand{\arXiv}[1]{\href{http://www.arXiv.org/abs/#1}{#1}}
\usepackage[colorlinks=true, linkcolor=blue, bookmarks=true]{hyperref}

\makeatletter
\renewcommand\section{\@startsection {section}{1}{\z@}%
                  {-3.5ex \@plus -1ex \@minus -.2ex}%nn
                  {2.3ex \@plus.2ex}%
                  {\normalfont\large\bfseries}}
\renewcommand\subsection{\@startsection{subsection}{2}{\z@}%
                   {-3.25ex\@plus -1ex \@minus -.2ex}%
                   {1.5ex \@plus .2ex}%
                   {\normalfont\bfseries}}
\makeatother

\newcommand{\beq}{\begin{equation}}
\newcommand{\eeq}{\end{equation}}
\newcommand{\ber}{\begin{array}}
\newcommand{\eer}{\end{array}}
\newcommand{\indices}{nlk}
\newcommand{\indo}{n_1l_1k_1}
\newcommand{\indt}{n_2l_2k_2}
\newcommand{\indth}{n_3l_3k_3}
\newcommand{\indf}{n_4l_4k_4}

\newcommand{\ci}{c_{nlk}}

\newcommand{\citb}{\bar{c}_{\indt}}
\newcommand{\cith}{c_{\indth}}
\newcommand{\cif}{c_{\indf}}
\newcommand{\cid}{\dot{c}_{\indices}}

\newcommand{\ciodd}{\ddot{c}_{\indo}}
\newcommand{\freqS}{(n+1)}
\newcommand{\freqSo}{(n_1+1)}
\newcommand{\del}{\partial}

\newcommand{\ei}{e_{\indices}}

\newcommand{\oi}{\omega_{\indices}}
\newcommand{\Y}{Y_{nlk}}
\newcommand{\ab}{\bar{a}}
\newcommand{\bb}{\bar{b}}
\newcommand{\yd}{\dot{y}}
\newcommand{\al}{\alpha}
\newcommand{\alb}{\bar{\alpha}}
\newcommand{\ald}{\dot{\alpha}}
\newcommand{\ali}{\alpha_{\indices}}
\newcommand{\alio}{\alpha_{\indo}}

\newcommand{\alitb}{\bar{\alpha}_{\indt}}
\newcommand{\alith}{\alpha_{\indth}}
\newcommand{\alif}{\alpha_{\indf}}

\newcommand{\aliod}{\dot{\alpha}_{\indo}}
\newcommand{\be}{\beta}
\newcommand{\beb}{\bar{\beta}}
\newcommand{\bed}{\dot{\beta}}
\newcommand{\bei}{\beta_{\indices}}
\newcommand{\beio}{\beta_{\indo}}

\newcommand{\beitb}{\bar{\beta}_{\indt}}
\newcommand{\beith}{\beta_{\indth}}
\newcommand{\beif}{\beta_{\indf}}

\newcommand{\beiod}{\dot{\beta}_{\indo}}
\newcommand{\B}[2]{B\left(#1,#2\right)}
\newcommand{\G}[1]{\Gamma{\left(#1\right)}}
\newcommand{\te}{\theta}
\newcommand{\ph}{\varphi}
\newcommand{\de}{\delta}
\newcommand{\De}{\Delta}

\newcommand{\eps}{\varepsilon}

\newcommand{\ena}{\end{eqnarray}}
\newcommand{\beqa}{\begin{eqnarray}}
\newcommand{\eeqa}{\end{eqnarray}}
\newcommand{\bea}{\begin{eqnarray}}
\newcommand{\eea}{\end{eqnarray}}

\newcommand{\shalf}{\frac{1}{2}}

\theoremstyle{remark}

\renewcommand{\Re}{\operatorname{Re}}
\renewcommand{\Im}{\operatorname{Im}}

\begin{document}
\begin{titlepage}
\begin{flushright}
\phantom{arXiv:yymm.nnnn}
\end{flushright}
\vspace{0cm}
\begin{center}
{\LARGE\bf Maximally rotating waves in AdS\vspace{2mm}\\ and on spheres}  \\
\vskip 15mm
{\large Ben Craps,$^{a}$ Oleg Evnin,$^{b,a}$  Vincent Luyten$^{a}$}
\vskip 7mm
{\em $^a$ Theoretische Natuurkunde, Vrije Universiteit Brussel (VUB) and\\
The International Solvay Institutes, Brussels, Belgium}
\vskip 3mm
{\em $^b$ Department of Physics, Faculty of Science, Chulalongkorn University,
Bangkok, Thailand}
\vskip 7mm
{\small\noindent {\tt Ben.Craps@vub.be, oleg.evnin@gmail.com, Vincent.Luyten@vub.be}}
\vskip 10mm
\end{center}
\vspace{2cm}
\begin{center}
{\bf ABSTRACT}\vspace{3mm}
\end{center}

We study the cubic wave equation in AdS$_{d+1}$ (and a closely related cubic wave equation on $S^3$) in a weakly nonlinear regime. Via time-averaging, these systems are accurately described by simplified infinite-dimensional quartic Hamiltonian systems, whose structure is mandated by the fully resonant spectrum of linearized perturbations. The maximally rotating sector, comprising only the modes of maximal angular momentum at each frequency level, consistently decouples in the weakly nonlinear regime. The Hamiltonian systems obtained by this decoupling display remarkable periodic return behaviors closely analogous to what has been demonstrated in recent literature for a few other related equations (the cubic Szeg\H o equation, the conformal flow, the LLL equation). This suggests a powerful underlying analytic structure, such as integrability. We comment on the connection of our considerations to the Gross-Pitaevskii equation for harmonically trapped Bose-Einstein condensates.

\vfill

%\begin{flushleft}
%PACS 11.25.-w, 04.65.+e
%\end{flushleft}
\end{titlepage}

%%%%%%%%%%%%%%%%%%%%%%%%%%%%%%%%%%%%%%%%%%%%%%%%%%%%
%%%%%%%%%%%%%%%%%%%%%%%%%%%%%%%%%%%%%%%%%%%%%%%%%%%%
%%%%%%%%%%%%%%%%%%%%%%%%%%%%%%%%%%%%%%%%%%%%%%%%%%%%

\section{Introduction}

Dynamics of nonlinear waves in confined domains is a fascinating subject, since (unlike in scattering situations) the interactions are never effectively cut off by the wave dispersal to infinity, resulting in sophisticated behaviors. Weakly turbulent phenomena, whereby nonlinearities transfer energy to progressively shorter wavelengths, are of particular interest in this setting.

When nonlinearities are introduced to systems whose linearized spectra of frequencies are perfectly resonant (all frequencies are commensurate), the sophisticated behaviors due to repeated wave scattering in the confined domain survive to arbitrarily small magnitudes of nonlinear interactions, provided that one waits long enough. (The transfer of energy between linearized normal modes becomes slow when the nonlinearities are weak.) A number of equations of mathematical physics display this feature. For instance, nonlinear dynamics of small perturbations of the Anti-de Sitter spacetime has attracted a lot of attention over the recent years (starting with \cite{BR}, for a review see \cite{review}). The same features are shared by nonlinear wave equations in AdS and on spheres, which will be the main subject of our present investigation. As we will show, these equations can also be viewed as a relativistic version of the Gross-Pitaevskii equation describing Bose-Einstein condensates in a harmonic trap (reviews can be found in \cite{BEC1,BEC2,BEC3}). This Gross-Pitaevskii equation also represents waves in a confined domain with a perfectly resonant spectrum of linear frequencies.

As a first step in analyzing nonlinear dynamics in confined domains for systems with resonant frequency spectra, it is natural to focus on weakly nonlinear regimes. 
%Perfectly resonant spectra allow sophisticated dynamics to survive to arbitrarily small nonlinearities, but this of course also makes the weakly nonlinear analysis highly nontrivial. 
Naive perturbative expansions break down due to so-called secular terms, and have to be replaced by alternative perturbative techniques. A number of such techniques are known, based on application of multiscale analysis, time-averaging or renormalization group resummation (for a textbook treatment, see \cite{murdock}, for discussions in the context of the AdS stability problem, see \cite{Balasubramanian:2014cja,CEV1,CEV2}).  As an output of these methods, one obtains a simplified infinite-dimensional {\it flow system} (also known as the resonant or effective system) describing slow energy transfer between the linearized modes due to resonant interactions.

Flow systems arising from the sort of weakly nonlinear analysis described above are often more structured than the original equations from which they descend, for example, they may have extra conservation laws \cite{BKS1,CEV2}. For some equations, the flow system may be simple enough to admit explicit analytic solution. For instance, both the conformal flow equation arising from the cubic wave equation on a 3-sphere, and the Lowest Landau Level (LLL) equation \cite{GHT,GT} arising from the Gross-Pitaevskii equation for harmonically trapped Bose-Einstein condensates, admit analytic solutions in which the linearized normal mode amplitudes exhibit exact periodic returns to the original configuration \cite{BCEHLM,ABCE}. Such remarkable solutions are likely to imply a deeper structure and allude to integrability. In fact, both of these flow equations look like more complicated generalizations of the cubic Szeg\H o equation, an integrable system designed in the mathematical literature \cite{GG} as a solvable model of weak turbulence.

In this article, our aim is to develop a series of flow systems arising from weakly nonlinear wave equations in AdS spacetime and on spheres. These flows generalize the conformal flow considered in \cite{BCEHLM} by treating wave equations without spherical symmetry, whereas the considerations of \cite{BCEHLM} were specific to perturbations on a 3-sphere rotationally invariant about one picked point. (We note that attempts to analyze similar problems for the considerably more involved related case of gravitational perturbations of AdS spacetime have recently intensified \cite{asymm1,asymm2,asymm3,asymm4}.) The flows we derive are also closely related to the LLL equation \cite{ABCE}, since they arise from focusing on maximally rotating modes (modes of maximal angular momentum from each frequency level), and there is furthermore a relation between the Gross-Pitaevskii equation and the wave equations we consider, as we explain in the discussion section. Remarkably, we find that the flow equations we derive display periodic perfect returns of the normal mode amplitude spectrum to the initial configuration, analogous to what has been known for the cubic Szeg\H o equation \cite{GG}, the conformal flow \cite{BCEHLM} and the LLL equation \cite{ABCE}.

While the systems we are considering are interesting in their own right from the nonlinear dynamics perspective, nonlinear wave equations in AdS spacetime have also surfaced in the context of gravitational holography research.
In string-theory-derived versions of the AdS/CFT correspondence, matter fields in the AdS bulk backreact on the metric, but this is not always the case in so-called bottom-up approaches to holography. For simplicity, one often starts by studying a regime in which the bulk matter fields of interest do not backreact on the bulk geometry. Such probe approximations have been used, for instance, in studies of holographic QCD \cite{Karch:2002sh, Sakai:2004cn} (in the limit in which the number of flavors is much smaller than the number of colors), holographic superconductors \cite{Hartnoll:2008vx, %Hartnoll:2008kx, 
Nishioka:2009zj} (in a limit in which the scalar operator that condenses has large charge), and holographic quantum quenches \cite{Basu:2011ft, Basu:2012gg} (in a limit in which a bulk scalar field has large self-coupling).

The paper is organized as follows. In section~\ref{sec:2}, we mainly study the conformally invariant cubic wave equation on the Einstein cylinder spacetime $R\times S^3$. We review the procedure of time-averaging, show that the resulting flow equations allow a consistent truncation to the maximally rotating sector, and construct exact solutions describing remarkable periodic returns. We also show that a similar attempt to construct periodic solutions fails for other dimensions. In section~\ref{sec:3}, we discuss the case of AdS, where the equations are a bit more involved, but where we are able to construct periodic return solutions in any dimension and for any value of the scalar field mass. In section~\ref{sec:4}, we conclude with a discussion of the significance of our results and possible further implications. In particular, we point out that the Gross-Pitaevskii equation with a harmonic potential can be viewed as a nonrelativistic limit of the cubic wave equation in AdS, which makes contact with a recent study of the LLL equation in \cite{ABCE}. 

%%%%%%%%%%%%%%%%%%%%%%%%%%%%%%%%%%%%%%%%%

\section{Weakly nonlinear dynamics of maximally rotating\\ perturbations on spheres}\label{sec:2}

We start by considering the wave equations on spatial spheres $S^d$, which correspond to the Einstein cylinder spacetime $R\times S^d$. We shall first specialize to $d=3$, the case for which the conformal flow of \cite{BCEHLM} was derived (in this dimension, the cubic wave equation enjoys symmetry enhancement to the full conformal group). We shall briefly comment on other dimensions at the end.

The relevant metric is (we set the 3-sphere radius to 1):
\beq \label{eq_metricS3}
ds^2=-dt^2+d\chi^2+\sin^2{\chi}(d\te^2+\sin^2{\te}\,d\ph^2).
\eeq
We choose to work with a complex scalar field $\phi$ for reasons that will become apparent below. The conformally invariant cubic wave equation is
\beq \label{eq_waveS3}
-\del_t^2\phi+\De_{S^3}\phi-\phi=|\phi|^2\phi,
\eeq
where $\De_{S^3}$ is the 3-sphere Laplacian given by
\beq
\De_{S^3}=\frac{1}{\sin^2{\chi}}\del_{\chi}\left(\sin^2{\chi}\del_{\chi}\right)+\frac{1}{\sin^2{\chi}}\De_{S^2},\qquad \De_{S^2}=\frac{1}{\sin{\te}}\del_{\te}\left(\sin{\te}\del_{\te}\right)+\frac{1}{\sin^2{\te}}\del_{\ph}^2.
\eeq
Note that if $\phi$ is replaced by a real field, this becomes identical to the wave equation treated in \cite{BCEHLM}.

The above wave equation can be converted into an infinite set of coupled oscillators by expanding $\phi$ in the basis of (hyper)spherical harmonics on $S^3$
\beq
\phi(t,\chi,\te,\ph)=\sum_{n=0}^{\infty}\sum_{l=0}^n\sum_{k=-l}^{l}\ci(t)\Y(\chi,\te,\ph),
\eeq
where the $\Y$ satisfy
\beq
\De_{S^3}\Y =-n(n+2)\Y ,
\eeq
and are normalised to one, such that \eqref{eq_waveS3} becomes
\beq \label{eq_oscS3}
\ciodd+\freqS^2c_{\indo} =-\hspace{-0.2cm}\sum_{\indt}\sum_{\indth}\sum_{\indf}C_{\indo\ldots\indf}\bar{c}_{\indt}c_{\indth}c_{\indf}.
\eeq
The interaction coefficients $C_{\indo\ldots\indf}$ are given by
\beq \label{eq_C_S3}
C_{\indo\ldots\indf}=\int d\Omega_3\bar{Y}_{\indo}\bar{Y}_{\indt}Y_{\indth}Y_{\indf},
\eeq
where $d\Omega_3=\sin^2\chi \sin\theta d\chi d\theta d\phi$ is the invariant measure on the 3-sphere. The hyperspherical harmonics on $S^3$ can be expressed in terms of the familiar spherical harmonics on $S^2$ by
\beq
Y_{\indices}\left(\chi,\te,\ph\right)=\sqrt{\frac{2(2l)!!(n+1)(n-l)!(2l+1)!}{\pi(2l+1)!!(n+l+1)!}}\,\sin^l{\chi}\,C^{(l+1)}_{n-l}\left(\cos{\chi}\right)\,Y_{lk}\left(\te,\ph\right),
\eeq
where the $C_n^{(\lambda)}(x)$ are Gegenbauer polynomials of degree $n$ with the measure parameter $\lambda$. They form a system of orthogonal polynomials on the interval $(-1,1)$ with respect to the measure $(1-x^2)^{\lambda-\shalf}$.

Note that the perfectly resonant spectrum of frequencies in (\ref{eq_oscS3}) is due to the conformal value of the mass in (\ref{eq_waveS3}) and would not be present for generic masses. (By contrast, in the AdS spacetimes we shall focus on in the next section, the spectrum is always fully resonant.) A fully resonant spectrum (more specifically, the property that the difference between any two frequencies is integer) is crucial to maintain the weakly nonlinear approximation in the form we are about to derive.

The solutions to the linearized system corresponding to  \eqref{eq_oscS3}, where the right-hand side has been replaced by zero, are simply
\beq
c_{\indices}^{\mbox{\tiny linear}}=A_{\indices}e^{i\freqS t}+B_{\indices}e^{-i\freqS t},
\eeq
where $A_{\indices}$ and $B_{\indices}$ are arbitrary complex constants.
One could then try to treat the non-linearity perturbatively by performing a weak field expansion of the form
\beq \label{eq_expphi}
\phi=\eps\phi_{\mbox{\tiny linear}}+\eps^3\phi_{\mbox{\tiny correction}}+\ldots,
\eeq
but $\phi_{\mbox{\tiny correction}}$ will grow in time, due to so called secular terms, and invalidate the above expansion at times of order $1/\eps^2$. This perturbative expansion can be resummed (improved), in different ways, leading to {\it flow equations} that accurately describe slow energy transfer between the modes due to nonlinearities, while the fast oscillations of the original linearized modes are `integrated out.' Rather than presenting this entire procedure, it is quicker (and equivalent) to directly factor out fast oscillations using a method known as time-averaging. (Some pedagogical comments on different approaches to improving perturbation theory can be found in \cite{CEV1,CEV2}.)

To employ time-averaging we first change variables from $c$ and $\dot c$  to the complex-valued functions of time $\ali(t)$ and $\bei(t)$
\begin{align}
\ci&=\eps\left(\ali e^{i\freqS t}+\bei e^{-i\freqS t}\right),\label{eq_cS3}\\
\cid&=i\eps\freqS(\ali e^{i\freqS t}-\bei e^{-i\freqS t}).\label{eq_cdotS3}
\end{align}
Combining these, we find for $\ali$
\beq
\eps\ali e^{i\freqS t}=\frac{1}{2}\left(\ci+\frac{\cid}{i(n+1)}\right).
\eeq
Differentiating and taking into account \eqref{eq_oscS3}, we find:
\beq \label{eq_preaveragingS3alpha}
2i\eps\freqSo\aliod=-\hspace{-2mm}\sum_{\indt}\sum_{\indth}\sum_{\indf}C_{\indo\ldots\indf}\citb\cith\cif e^{-i\freqSo t}.
\eeq
Analogously,
\beq \label{eq_preaveragingS3beta}
2i\eps\freqSo\beiod=\hspace{-2mm}\sum_{\indt}\sum_{\indth}\sum_{\indf}C_{\indo\ldots\indf}\citb\cith\cif e^{i\freqSo t}.
\eeq
In the above equations, the $\ci$ should be expressed through $\ali$ and $\bei$. This leads to a collection of terms on the right-hand side oscillating as $e^{-i\Omega t}$, where $\Omega$ is $(n_1+1)\pm (n_2+1)\pm (n_3+1)\pm (n_4+1)$ for (\ref{eq_preaveragingS3alpha}) and $-(n_1+1)\pm (n_2+1)\pm (n_3+1)\pm (n_4+1)$ for (\ref{eq_preaveragingS3beta}). All the plus-minus signs are independent. We call the terms with $\Omega=0$ resonant interactions and those with $\Omega\neq 0$ non-resonant.

Rewriting (\ref{eq_preaveragingS3alpha}-\ref{eq_preaveragingS3beta}) in  terms of the slow time $\tau=\eps^2t$, the dependence on $\eps$ drops out, except that the non-resonant terms are now proportional to $e^{-i\Omega\tau/\eps^2}$. This means that in the weak field regime, $\eps \rightarrow 0$, they become highly oscillatory and time-averaging is equivalent to simply discarding all non-resonant terms. It can be proved that the resulting time-averaged system with discarded non-resonant terms accurately approximates the original system on time scales of order $1/\eps^2$ for small $\eps$ \cite{murdock}.

After the non-resonant terms in (\ref{eq_preaveragingS3alpha}-\ref{eq_preaveragingS3beta}) have been discarded, one further simplification occurs. There are solutions to the resonance condition $\Omega=0$ of the form $n_1=2+n_2+n_3+n_4$ (or other variants obtained by permuting the $n$'s). However, all the interaction coefficients $C$ for such terms will vanish. This is completely analogous to the usual angular momentum selection rules in quantum mechanics. The integral in (\ref{eq_C_S3}) is over the invariant measure on the sphere, and it can only be nonzero if the direct product of the representations of the sphere isometries furnished by the four factors in the integrand contains the identity representation. The spherical harmonics are in the rank $n$ traceless fully symmetric tensor representations. It is impossible to contract four traceless fully symmetric tensors of ranks $n_1$, $n_2$, $n_3$, $n_4$ satisfying $n_1=2+n_2+n_3+n_4$ to obtain a scalar. Therefore, the corresponding $C$-coefficients vanish. The only contributing solutions to the resonance condition $\Omega=0$ are thus of the form $n_1+n_2=n_3+n_4$ (and permutations of the $n$'s). Such selections rules are well-known for analogous considerations in AdS \cite{CEV1,CEV2,Yang,EK,EN}.

Putting everything together, and taking into account the index permutation symmetries of the coefficients $C$ given by (\ref{eq_C_S3}), we obtain the following time-averaged equations for $\alpha$:
\beq \label{eq_preflowS3alpha}
\begin{split}
2i\freqS\frac{d\alio}{d\tau}=&-\underset{n_1+n_2=n_3+n_4}{\sum_{\indt}\sum_{\indth}\sum_{\indf}}C_{\indo\ldots\indf}\alitb\alith\alif \\
&-2\underset{n_1+n_3=n_2+n_4}{\sum_{\indt}\sum_{\indth}\sum_{\indf}}C_{\indo\ldots\indf}\beitb\beith\alif.
\end{split}
\eeq
As we remarked, this equation accurately approximates the original system on time scales $\mathcal{O}(\eps^{-2})$, in the sense that the difference of exact solutions and solutions obtained from the time-averaged system uniformly becomes arbitrarily small on such intervals for small $\eps$ \cite{murdock}.
The time-averaged system for $\beta$ becomes
\beq \label{eq_preflowS3beta}
\begin{split}
2i\freqS\frac{d\beio}{d\tau}=&-\underset{n_1+n_2=n_3+n_4}{\sum_{\indt}\sum_{\indth}\sum_{\indf}}C_{\indo\ldots\indf}\beitb\beith\beif \\
&-2\underset{n_1+n_3=n_2+n_4}{\sum_{\indt}\sum_{\indth}\sum_{\indf}}C_{\indo\ldots\indf}\alitb\alith\beif.
\end{split}
\eeq

We observe that time-averaging has enhanced the symmetry. While the original system possesed a $U(1)$ symmetry rotating $\alpha$ and $\beta$ by the same common phase (the usual $U(1)$ symmetry of a charged complex scalar), the time-averaged system allows rotation of all $\alpha$'s and all $\beta$'s by two independent common phases, thus giving two $U(1)$ groups. This is closely related to the apearance of a new $U(1)$ symmetry in the time-averaged system for real scalar fields described in the literature \cite{BKS1,CEV2} and resulting in extra conservation laws. In close relation to this, $\beta$ can be consistently set to zero in the time-averaged equations, resulting in the following system containing only $\alpha$: 
\beq \label{eq_flowS3}
i\freqS\aliod=\underset{n_1+n_2=n_3+n_4}{\sum_{\indt}\sum_{\indth}\sum_{\indf}}C_{\indo\ldots\indf}\alitb\alith\alif.
\eeq
(We have rescaled $\tau$ to eliminate numerical factors.) We shall henceforth focus on the sector described by this equation (and disregard the $\beta$-variables).

Equation (\ref{eq_flowS3}) is still rather complicated, and a reasonable strategy is to look for smaller decoupling subsectors, which can be analyzed independently. One example is the spherically symmetric scalar field, which amounts to retaining only the modes with $l=k=0$. The resulting equation is the conformal flow, studied previously in \cite{BCEHLM} and arising there from a real scalar field wave equation. In  \cite{BCEHLM}, a range of remarkable properties were demonstrated for the spherically symmetric truncation of (\ref{eq_flowS3}), including explicit analytic solutions for which $|\alpha(t)|$ is a periodic function of time with a common period for all modes. Our main purpose of this article is to demonstrate that a few distinct systems sharing the same property emerge from other consistent truncations of (\ref{eq_flowS3}) and other related equations.

The specific sector of (\ref{eq_flowS3}) we want to focus on can be called the maximally rotating sector. These are the modes with $n=l=k$, i.e., the modes of maximal angular momentum for each frequency. The decoupling can be seen as follows. First note that the mode functions are all proportional to $e^{-ik\ph}$, where $k$ is the number of units of angular momentum along the polar axis. All three integrals in the interaction coefficients (\ref{eq_C_S3}) factorize, so that the integral over $\ph$ ensures that only coefficients $k_1+k_2=k_3+k_4$ are non-zero (this is just the angular momentum conservation in mode interactions). 
Consider the situation where the only excited modes are maximally rotating. Then the time-derivatives of the non-maximally rotating modes are zero, as can be seen from \eqref{eq_flowS3}: only terms with coefficients $n_2=k_2$,  $n_3=k_3$, $n_4=k_4$ are non-zero, but the summation restriction and properties of $C$ then impose $n_1+n_2=n_3+n_4$ and $k_1+n_2=n_3+n_4$, so that that $k_1$ (and hence $l_1$) equal $n_1$. Therefore, non-maximally rotating modes are never excited. In the rest of this section we focus on the dynamics in the maximally rotating subsector. (Note that restriction to the maximally rotating sector is incompatible with reality of the field $\phi$. That is the reason why we had to start with a complex field.)

The maximally rotating spherical harmonics are given by
\beq \label{eq_maxspinmodeS3}
e_n(\chi,\te,\ph)=Y_{nnn}(\chi,\te,\ph)=\frac{\sqrt{n+1}}{\sqrt{2}\pi}\sin^n{\chi}\sin^n{\te}e^{-in\ph}.
\eeq
Correspondingly the interaction coefficients are evaluated as (we discard mode number independent numerical factors as they can be absorbed in a rescaling of time):
\beq
C_{nmkl}=\int d\Omega_3 e_ne_me_ke_l =\frac{\sqrt{(n+1)(m+1)(k+1)(l+1)}}{n+m+1}.
\eeq
The `maximally rotating flow' equation is then
\beq \label{eq_maxspinflow}
i(n+1)\dot{\alpha}_n=\sum_{j=0}^{\infty}\sum_{k=0}^{n+j}\frac{\sqrt{(n+1)(j+1)(k+1)(n+j-k+1)}}{n+j+1}\alb_j\al_k\al_{n+j-k}.
\eeq
In addition to the scaling symmetry 
$\alpha \rightarrow \lambda \alpha(\tau/\lambda^2)$
, this equation possesses two further symmetries (where $\xi_1$ and $\xi_2$ are real parameters),
\beq
\ali \rightarrow e^{i\xi_1}\ali,\qquad \ali \rightarrow e^{in\xi_2}\ali,
\eeq
giving rise to two conserved quantities
\beq
Q=\sum_{n=0}^{\infty}(n+1)|\alpha_n|^2,\qquad
E=\sum_{n=0}^{\infty}(n+1)^2|\alpha_n|^2.
\eeq
To simplify (\ref{eq_maxspinflow}), introduce
\beq
\beta_n\equiv \sqrt{n+1}\,\alpha_n
\eeq
(unrelated to the modes $\beta_{nlk}$ appearing at early stages of our derivations that we have consistently set to zero) and rewrite it as
\beq \label{eq_maxspinflowbeta}
i\dot{\beta}_n=\sum_{j=0}^{\infty}\sum_{k=0}^{n+j}\,\frac{\bar\beta_j\beta_k\beta_{n+j-k}}{n+j+1}.
\eeq

Analogously to the (spherically symmetric) conformal flow of \cite{BCEHLM}, we try to find low-dimensional invariant submanifolds of (\ref{eq_maxspinflowbeta}) by making an ansatz depending on only a few parameters and seeing whether it closes. Motivated by the spherically symmetric case, we choose
\beq\label{s3ansatz}
\beta_n=(b+an)p^n,
\eeq
where $a$, $b$ and $p$ are complex-valued functions of time. Substitution in \eqref{eq_maxspinflowbeta} gives
\beq\label{substitution}
i\left(\dot{b}+n\left(\dot{a}+b\frac{\dot{p}}{p}\right)+n^2a\frac{\dot{p}}{p}\right)=\sum_{j=0}^{\infty}\frac{(\ab+j\bb)|p|^{2j}}{n+j+1}\sum_{k=0}^{n+j}(b+ka)(b+(n+j-k)a).
\eeq
Note that the factor of $p^n$ has consistently cancelled between the two sides.

We will have a consistent ansatz if the right-hand side of (\ref{substitution}) is a quadratic polynomial in $n$ after the summations have been performed. We first use
\beq
\sum_{k=0}^{N}k = \frac{N(N+1)}{2}, \qquad
\sum_{k=0}^{N}k^2 = \frac{N(N+1)(2N+1)}{6}.
\eeq
Note that these Faulhaber's sums are divisible by $N+1$, so that the factor of $1/(n+j+1)$ cancels, leaving at most quadratic terms in $n$. This guarantees the closure of our ansatz. 
It remains to carry out the sums over $j$ using
\begin{align}
\sum_{j=0}^{\infty}j^Ax^j&=(x\del_x)^A\frac{1}{1-x}.\label{eq_geomsum}
\end{align}

Setting the coefficients of equal powers of $n$ equal we arrive at a system of three equations for three variables, which can be conveniently rewritten as
\begin{align}
\frac{i\dot{a}}{1+y} &= -\frac{a^2\bb}{6}+\frac{5a|b|^2}{6}+y\left(\frac{5|a|^2b}{6}+\frac{a|a|^2}{6}+\frac{a^2\bb}{3}\right)+2y^2\frac{a|a|^2}{3}, \label{eq_adot}\\
\frac{i\dot{b}}{1+y} &= b|b|^2+y\left(\ab b^2+a|b|^2+\frac{a^2\bb}{3}+|a|^2b-\frac{a|a|^2}{6}+\frac{b|b|^2}{6}\right)\nonumber\\
&\quad +2y^2\left(|a|^2b-\frac{a|a|^2}{6}+\frac{a^2\bb}{6}+3\frac{b|b|^2}{6}\right)+6y^3\frac{b|b|^2}{6},\label{eq_bdot}\\
\frac{i\dot{p}}{1+y} &=\frac{p}{6}\left(a\bb + y|a|^2\right) \label{eq_pdot},
\end{align}
where we have introduced
\beq \label{ydef}
y=\frac{|p|^2}{1-|p|^2}.
\eeq

We will now explicitly solve the dynamics on this invariant subspace. Equation \eqref{eq_pdot} can be converted into an equation for $y$:
\beq \label{eq_ydot}
\yd=\frac{1}{3}y(1+y)^2\Im{(a\bb)}.
\eeq
Using conservation laws, we will be able to reduce the system to a single equation for $y$. With the sums \eqref{eq_geomsum}, the conserved quantities $Q$ and $E$ take the following form in terms of the parameters of our ansatz:
\begin{align}
Q&=(1+y)\left[|b|^2+2\Re{(a\bb)}y+|a|^2y(1+2y)\right], \label{eq_Qansatz}\\
E&=(1+y)^2\left[|b|^2+4\Re{(a\bb)}y+2|a|^2y(1+3y)\right].\label{eq_Eansatz}
\end{align}
The Hamiltonian gives the third independent conservation law, but it is more convenient to derive another related conserved quantity quadratic in $a$. We first write
\beq
\frac{d|a|^2}{d\tau}=-\frac{1}{3}(1+y)(1+3y)\Im{(a\bb)}.
\eeq
Combined with \eqref{eq_ydot}, this ensures the conservation of
\beq \label{eq_S}
S=|a|^2y(1+y)^2.
\eeq

Expressing $|b|^2,|a|^2$ and $\Re{(a\bb)}$ through $Q,E,S$ and $y$ we get
\begin{align}
|a|^2&=\frac{S}{y(1+y)^2},\\\label{aabs}
\Re{(a\bb)}&=\frac{E-Q(1+y)-S(1+4y)}{2y(1+y)^2},\\
|b|^2&=\frac{2Q(1+y)-E+2Sy}{(1+y)^2}.\label{babs}
\end{align}

Inserting these expressions in \eqref{eq_ydot} we find the following equation for $y$
\beq
\yd^2=\frac{1}{36}\left(-(E-Q-S)^2+2((2E-Q-4S)S+(E-Q)Q)y-(Q^2+8S^2)y^2\right),
\eeq
which expresses the energy conservation for an ordinary one-dimensional harmonic oscillator. This immediately guarantees that all solutions for $y$, and hence for $|p|$ are exactly periodic with period
\beq
T=\frac{12\pi}{\sqrt{Q^2+8S^2}}.
\eeq
Equations (\ref{aabs}-\ref{babs}) then guarantee that the same property is shared by $|a|^2$, $|b|^2$ and $\Re(a\bar b)$. From this it follows that the absolute values of the amplitudes $|\alpha_n|$ are exactly periodic with the same common period. We have thus recovered in our maximally rotating sector the periodic behaviors observed in the literature for the conformal flow, the cubic Szeg\H o equation and the LLL equation.

We briefly comment on what happens when one tries to generalize the above story to general dimensions $d$. We will denote the angles on $S^d$ as $\te_1,\ldots\te_{d-1},\ph$ and collectively as $\Omega$. The metric is most compactly expressed recursively in terms of metrics on lower-dimensional spheres,
\beq
d\Omega_d^2=(d\te_{d-1})^2+\sin^2\te_dd\Omega_{d-1}^2,
\eeq
resulting in a recursion relation for the Laplacian
\beq
\Delta_{S^{d}}=\frac{1}{\sin^{d-1}\te_{d-1}}\del_{\te_{d-1}}\left(\sin^{d-1}\te_{d-1}\del_{\te_{d-1}}\right)+\frac{1}{\sin^2\te_{d-1}}\Delta_{S^{d-1}}.
\eeq

For general $d$, the mass term in the wave equation required to ensure a fully resonant linearized spectrum is $-(d-1)^2\phi/4$, corresponding to the conformal mass. (Note that this reduces to the previous $-\phi$ for $d=3$.) The wave equation is then given by
\beq \label{eq_waveSd}
-\del_t^2\phi+\Delta_{S^d}\phi-\frac{(d-1)^2}{4}\phi=|\phi|^2\phi,
\eeq
and hence the mode functions are now spherical harmonics on $S^d$, which also have a corresponding recursive expression
\beq
Y_{nl\mu}\left(\te_{d-1},\ldots,\te_1,\ph\right)=N_{nl}\sin^l\te_{d_1}C^{(l+\frac{d}{2}-1)}_{n-l}\left(\cos\te_{d-1}\right)Y_{l\mu}\left(\te_{d-2},\ldots,\te_1,\ph\right),
\eeq
 with an appropriate normalization factor $N_{nl}$.
We use $\mu$ as a shorthand for all the other indices present in the successively lower-dimensional harmonics. The spherical harmonics are eigenfunctions of the Laplacian on $S^d$ with eigenvalues $-n(n+d-1)$. Hence, the corresponding oscillation frequencies are $n+(d-1)/2$, giving a perfectly resonant spectrum. The maximally rotating harmonics are given by
\beq
e_n(\te_1,\ldots\te_{d-1},\ph)=\sqrt{\frac{\G{n+\frac{d+1}{2}}}{\G{n+1}}}\sin^n{\te_1}\ldots\sin^n{\te_{d-1}}e^{-in\ph}.
\eeq

Repeating the time-averaging procedure with the generalized expressions for the mass and eigenvalues, we find the maximally rotating flow equation for general $d$
\beq \label{eq_maxspinflowSd}
i\left(n+\frac{d-1}{2}\right)\dot{\alpha}_n=\sum_{j=0}^{\infty}\sum_{k=0}^{n+j}C_{njkn+j-k}\alb_j\al_k\al_{n+j-k},
\eeq
where the interaction coefficients are given by (again up to irrelevant numerical factors)
\beq\label{sdc}
\textstyle C_{nmkl}=\int d\Omega_d e_ne_me_ke_l =\sqrt{\frac{\G{n+\frac{d+1}{2}}}{\G{n+1}}\frac{\G{m+\frac{d+1}{2}}}{\G{m+1}}\frac{\G{k+\frac{d+1}{2}}}{\G{k+1}}\frac{\G{l+\frac{d+1}{2}}}{\G{l+1}}}\frac{\G{n+m+1}}{\G{n+m+\frac{d+1}{2}}}.
\eeq
Here $d\Omega_d=d\ph\Pi_{l=1}^{d-1}\sin^{d-l}\te_l d\te_l$ is the integration measure on $S^d$. 

The  flow equation on $S^d$ generalizes the case of $S^3$ analyzed above, and is also extremely similar to the flow equations in AdS treated in the next section. These parallels suggest that one should try the ansatz
\beq
\be_n\equiv\left(n+\frac{d-1}{2}\right)\sqrt{\frac{\G{n+1}}{\G{n+\frac{d+1}{2}}}}\,\al_n=(b+an)p^n,
\eeq
which reduces to (\ref{s3ansatz}) at $d=3$ and simplifies the equations (in particular, Faulhaber's sums again make an appearance). One discovers, however, that evaluation of the sums does not produce the necessary polynomial dependence on $n$, and the ansatz thus fails. We conclude that it is unlikely that the weak field dynamics on $S^d$ displays an invariant manifold analogous to what we have seen on $S^3$ and what we are about to see for AdS of any dimension in the next section.

%%%%%%%%%%%%%%%%%%%%%%%%%%%%%%%%%%%%%%%%%

\section{Weakly nonlinear dynamics of maximally rotating\\ perturbations in AdS}\label{sec:3}

We now turn to the case of global Anti-de Sitter spacetime AdS$_{d+1}$ with $d$ spatial dimensions, and consider a complex scalar field with a general mass $m$. AdS spacetime is remarkable in that the spectrum of frequencies of linear fields is fully resonant for any mass in any dimension (which has a simple explanation in terms of the algebra of AdS isometries). The AdS metric with radius set to $1$ is
\beq
ds^2=\frac{1}{\cos^2{x}}\left(-dt^2+dx^2+\sin^2{x}\,d\Omega_{d-1}^2\right),
\eeq
where $d\Omega_{d-1}^2$ is the metric on the $(d-1)$-sphere parametrized in hyperspherical coordinates collectively denoted as $\Omega$. On this spacetime, the complex scalar wave equation for a field of mass $m$ with a cubic non-linearity is
\beq\label{AdSwave}
\cos^2{x}\left(-\del_t^2\phi+\frac{1}{\tan^{d-1}{x}}\del_x\left(\tan^{d-1}{x}\del_x\phi\right)+\frac{1}{\sin^2{x}}\De_{S^{d-1}}\phi\right)-m^2\phi=|\phi|^2\phi,
\eeq
where $\De_{S^{d-1}}$ is the $(d-1)$-sphere Laplacian. The linearized system can be solved by separation of variables. First one computes the mode functions as solutions of the eigenvalue problem
\beq \label{eq_eigenvalueAdS}
\left(\frac{1}{\tan^{d-1}{x}}\del_x\left(\tan^{d-1}{x}\del_x\right)+\frac{1}{\sin^2{x}}\De_{\Omega_{d-1}}-\frac{m^2}{\cos^2{x}}\right)\ei(x,\Omega)=-\oi^2\ei(x,\Omega).
\eeq
The expansion of $\phi$ in these modefunctions then yields the general linearized solution
\beq\label{AdSlin}
\phi_{\mbox{\tiny linear}}(t,x,\Omega)=\sum_{n=0}^{\infty}\sum_{l,k}(A_{\indices}e^{i\oi t}+B_{\indices}e^{-i\oi t})\ei(t,x,\Omega),
\eeq
where $A_{\indices}$ and $B_{\indices}$ are arbitrary complex constants.

The explicit form of the mode functions is known as
\beq \label{eq_modefunctionAdS}
\ei(x,\Omega)=\mathcal{N}_{\indices}\cos^{\de}{x}\sin^l{x}P_n^{\left(\de-\frac{d}{2}, l+\frac{d}{2}-1\right)}(-\cos{2x})Y_{lk}(\Omega)
\eeq
and
\beq \label{eq_eigenvalueAdS}
\oi=\de+2n+l,
\eeq
where $\de=\frac{d}{2}+\sqrt{\frac{d^2}{4}+m^2}$ and $\mathcal{N}_{\indices}$ is a normalisation factor. (Note that the difference of any two frequencies is integer irrespectively of $\de$.) The $P_n^{(a,b)}(x)$ are the Jacobi polynomials, an orthogonal basis on the interval $(-1,1)$ with respect to the measure $(1-x)^a(1+x)^b$. The $Y_{lk}$ are spherical harmonics in $(d-1)$ dimensions, i.e.\ eigenfunctions of the corresponding sphere Laplacian with eigenvalue $l(l+d-2)$, and $k$ labels all harmonics contained in a given $l$-multiplet.

One can perform a weakly nonlinear analysis of (\ref{AdSwave}) in a manner exactly identical to the previous section. After implementing time-averaging, one obtains a system of flow equations describing slow evolution of the complex amplitudes $\alpha_{nlk}(t)$ and $\beta_{nlk}(t)$ descending from the constant amplitudes $A_{nlk}$ and $B_{nlk}$ in the linearized solution (\ref{AdSlin}). Due to selection rules in the interaction coefficients \cite{CEV1,CEV2,Yang,EK,EN}, the flow equations enjoy enhanced symmetries that permit consistently setting all $\beta$'s to zero. Furthermore, the resulting equation for $\alpha$ can be consistently truncated to the maximally rotating sector, comprising modes of maximal angular momentum at each frequency level (this is a consequence of the resonance condition on frequencies of the interacting modes and angular momentum conservation). In the notation of \eqref{eq_eigenvalueAdS}, maximally rotating modes exactly correspond to $n=0$. The modes we retain are then labelled by a single number, the polar axis projection of their angular momentum $m$, and are denoted simply by $\alpha_m$. We thus arrive at the maximally rotating conformal flow equation on AdS$_{d+1}$
\beq
i(\de+n)\ald_n=\sum_{m=0}^{\infty}\sum_{k=0}^{n+m}C_{nmk,n+m-k}\alb_m\al_k\al_{n+m-k},
\eeq
where the interaction coefficients are given by
\beq
C_{nmjk}=\int_0^{\frac{\pi}{2}}dx\frac{\tan^{d-1}{x}}{\cos^2{x}} \int d\Omega_{d-1} e_ne_me_je_k.
\eeq
Here, $d\Omega_{d-1}$ is the integration measure on $S^{d-1}$ given below (\ref{sdc}).
This equation possesses the same symmetries as \eqref{eq_flowS3} and hence the corresponding conserved quantities
\beq\label{AdSconserved}
Q=\frac{1}{\G{\de}}\sum_{n=0}^{\infty}(n+\de)|\alpha_n|^2,\qquad
E=\frac{1}{\G{\de}}\sum_{n=0}^{\infty}(n+\de)^2|\alpha_n|^2,
\eeq
where we have divided by $\G{\de}$ for future convenience.

The maximally rotating modes are given by (again, we omit plain numerical factors independent of the mode number, as they can always be absorbed in a redefinition of time):
\beq
e_n(x,\te_1,\dots, \te_{d-2},\ph)\equiv e_{0n\cdots n}(x,\Omega)=\sqrt{\frac{\G{n+1+\de}}{\G{n+1}}}\cos^{\de}{x}\sin^n{\te_1}\dots\sin^n{\te_{d-2}}e^{-in\ph},
\eeq
where we have written out explicitly the angles $\te_1,\ldots\te_{d-2},\phi$ collectively denoted by $\Omega$.
The interaction coefficients can be evaluated as
\beq
C_{nmjk}= 
 \sqrt{\frac{\G{n+1+\de}\G{m+1+\de}\G{j+1+\de}\G{k+1+\de}}{\G{n+1}\G{m+1}\G{j+1}\G{k+1}}}\frac{\G{n+m+1}}{\G{n+m+2\de}}.
\eeq
(This expression is nearly identical to the formula on $S^d$ from the previous section, but the minor difference will play a crucial role in our subsequent derivation.) Note that at this point, the number of AdS dimensions and the scalar field mass only enter the equations through $\de$ (which is also known as the `conformal dimension').

Once again, we try to find a finite-dimensional dynamically invariant subspace. To this end, define
\beq \label{eq_defbeta}
\beta_n\equiv (n+\de)\sqrt{\frac{\G{n+1}}{\G{n+1+\de}}}\al_n,
\eeq
in terms of which the flow equation in AdS becomes
\beq \label{eq_adsflow}
i\bed_n=\sum_{m=0}^{\infty}\frac{\G{m+\de}}{\G{m+1}}\sum_{k=0}^{n+m}\binom{n+m}{k}\B{k+\de}{n+m-k+\de}\beb_m\be_k\be_{n+m-k}.
\eeq

Analogously to the considerations of the previous section, we examine the ansatz
\beq \label{eq_adsansatz}
\be_n=(b+na)p^n
\eeq
to see if it is respected by the evolution.
The sum over $k$ in (\ref{eq_adsflow}) can be computed as follows. First, for the sum without powers of $k$,
\beq
\begin{aligned}
\sum_{k=0}^{N}\binom{N}{k}\B{k+\de}{N-k+\de}&=\int_0^1dx\sum_{k=0}^N\binom{N}{k}x^{k+\de-1}(1-x)^{N-k+\de-1}\\
&=\int_0^1dx\ x^{\de-1}(1-x)^{\de-1}=\B{\de}{\de}.
\end{aligned}
\eeq
The sums involving powers of k are analogously computed as
\begin{align}
\sum_{k=0}^{N}k\binom{N}{k}\B{k+\de}{N-k+\de}&=N\B{\de+1}{\de}=\frac{N}{2}\B{\de}{\de},\\
\sum_{k=0}^{N}k^2\binom{N}{k}\B{k+\de}{N-k+\de}
&=\left(\frac{N}{2}+\frac{N(N-1)}{2}\frac{1+\de}{1+2\de}\right)\B{\de}{\de}.
\end{align}

Finally, one carries out the $m$-summation using
\beq
\sum_{m=0}^{\infty}\frac{\G{m+\de}}{\G{m+1}}m^Ax^m=(x\del_x)^A\frac{\G{\de}}{(1-x)^{\de}}.
\eeq
At the end of the day, one obtains quadratic polynomials in $n$ on both sides of  (\ref{eq_adsflow}) , which ascertains the validity of the ansatz and results in the following equations:
\begin{align}
\frac{i\dot{b}}{(y+1)^{\de}}&=b|b|^2+y\de\left(\ab b^2+a|b|^2+|a|^2b\right)+\label{eq_bdot_AdS}\\
& \quad y^2\de\left(|a|^2b(\de+1)+\frac{a^2\bb}{2}\frac{\de(\de+1)}{1+2\de}+a|a|^2\frac{\de(\de+1)}{1+2\de}\right)+y^3\frac{a|a|^2}{2}\frac{\de^2(\de+1)(\de+2)}{1+2\de}, \nonumber \\
\frac{i\dot{a}}{(y+1)^{\de}}&=\frac{a|b|^2}{2}\frac{2+3\de}{1+2\de}-\frac{a^2\bb}{2}\frac{\de}{1+2\de}+y\de\left(\frac{|a|^2b}{2}\frac{2+3\de}{1+2\de}+a^2\bb\frac{\de}{1+2\de}+\frac{a|a|^2}{2}\frac{\de}{1+2\de}\right)+\nonumber\\
& \quad y^2a|a|^2\frac{\de^2(\de+1)}{1+2\de},\label{eq_adot_AdS}\\
\frac{i\dot{p}}{(y+1)^{\de}}&=\frac{p}{2}\de\left(a\bb\frac{1}{1+2\de}+y|a|^2\frac{\de}{1+2\de}\right).\label{eq_pdot_AdS}
\end{align}
We have absorbed an overall factor of $B(\de,\de)\G{\de}$ in another rescaling of time.

We will solve these equations in a manner completely analogous to the maximally rotating flow of the previous section. 
Expressing the conserved quantities (\ref{AdSconserved}) within our ansatz yields
\begin{align}
Q&=(y+1)^{\de}\left[|b|^2+2\Re{(a\bb)}\de y+|a|^2\de y(1+(\de+1)y)\right],\\
E&=\de (y+1)^{\de+1}\left[|b|^2+2\Re{(a\bb)}(\de+1) y+|a|^2(\de+1)y(1+(\de+2)y)\right].
\end{align}
We then convert the equation for $p$ into one for $y$, as defined by (\ref{ydef}):
\beq \label{eq_ydot_AdS}
\frac{\yd}{(y+1)^{\de+1}}=\frac{y\de}{1+2\de}\Im{(a\bb)},
\eeq
and obtain an extra conserved quantity $S$ from this equation and the expression for the time derivative of $|a|^2$ derived from \eqref{eq_adot_AdS}:
\beq
S=|a|^2y(y+1)^{\de+1}.
\eeq
Expressing $|b|^2$, $|a|^2$ and $\Re{(a\bb)}$ through $Q$, $E$, $S$ and  $y$ we find
\begin{align}
|a|^2&=\frac{S}{y(y+1)^{\de+1}}, \label{adsa2}\\
\Re{(a\bb)}&=\frac{E}{2\de y(y+1)^{\de+1}}-\frac{Q}{2y(y+1)^{\de}}-\frac{(\de+1)S}{y^{\de+1}}-\frac{S}{2y(y+1)^{\de+1}},\\
|b|^2&=\de(\de+1)\frac{y}{(y+1)^{\de+1}}S-\frac{E}{(y+1)^{\de+1}}+(\de+1)\frac{Q}{(y+1)^{\de}}.\label{adsb2}
\end{align}
Inserting this in \eqref{eq_ydot_AdS} we arrive at
\beq
\begin{split}
\yd^2=&-y^2\frac{\de^2}{4(1+2\de)^2}\left(Q^2+4(\de+1)S^2\right)\\
& -y\frac{\de}{2(1+2\de)^2}\left(-E(Q+2(1+\de)S)+\de(Q^2+(Q+2E)S+2(1+\de)S^2)\right)\\
&-\frac{(E-(Q+S)\de)^2}{4(1+2\de)^2},
\end{split}
\eeq
which again expresses energy conservation of a harmonic oscillator. This immediately guarantees that all solutions for $y$, and hence for $|p|$ are exactly periodic with period
\beq
T=\frac{4\pi(1+2\de)}{\de\sqrt{Q^2+4(\de+1)S^2}}.
\eeq
Equations (\ref{adsa2}-\ref{adsb2}) then guarantee that the same property is shared by $|a|^2$, $|b|^2$ and $\Re(a\bar b)$. From this it follows that the absolute values of the amplitudes $|\alpha_n|$ are exactly periodic with the same common period. Maximally rotating flows in AdS thus share the same periodic return property previously described for the conformal flow, the cubic Szeg\H o equation, the LLL equation and the maximally rotating flow on $S^3$.

%%%%%%%%%%%%%%%%%%%%%%%%%%%%%%%%%%%%%%%%%

\section{Discussion}\label{sec:4}

We have considered cubic wave equations for a complex scalar field on Einstein static universes and in AdS spacetimes of various dimensions. In all cases considered, the spectrum of frequencies of linear normal modes is perfectly resonant (the difference of any two frequencies is an integer). Nontrivial effects of nonlinearities survive to arbitrarily small field amplitudes and can be effectively described by simplified flow systems capturing the slow energy transfer between the normal modes in weakly nonlinear regimes. These flow systems can be consistently truncated to maximally rotating modes (only one mode carrying the maximal angular momentum is retained from each frequency level). This is analogous to the Lowest Landau Level (LLL) truncation of the Gross-Pitaevskii equation describing harmonically trapped Bose-Einstein condensates. The resulting maximally rotating flow systems appear to be highly structured analytically and admit simple explicit analytic solutions with exactly periodic energy flows for $S^3$ and in AdS of any dimension. (Such explicit solutions are extraordinary for infinite-dimensional nonlinear systems and allude at deeper and more far-reaching structures, such as integrability.) We shall now briefly comment on the implications of our findings.

The weakly nonlinear dynamics of a cubic conformal wave equation on $R\times S^3$ has been previously treated in \cite{BCEHLM}, where the analysis was restricted to the spherically symmetric sector and the weak field dynamics displayed periodic behaviors of the same type we found here. (This, in fact, was among the main motivations for our present work.) That the same equation for a complex scalar field, now truncated to the completely different maximally rotating sector, displays similar analytic structures, makes us strongly suspect that the full weakly nonlinear dynamics of the cubic conformal wave equation on $R\times S^3$, without any mode truncation, is analytically tractable. We leave this subject for future work. (Note that because of the conformal relation between $R\times S^3$ and AdS$_4$, our considerations of the maximally rotating sector in AdS$_4$ provide yet another sector of $S^3$ dynamics decoupling in the weak field regime and displaying exact returns of the energy spectrum. The cubic wave equation on $R\times S^d$ for $d\ne 3$ is not conformally invariant and thus cannot be mapped to a cubic wave equations in AdS.)

We have not succeeded in obtaining similar returning solutions on $R\times S^d$ with $d\ne 3$, and strongly suspect that the dynamical features at $d=3$ are not shared by the spheres of other dimensions. By contrast, we see returning behaviors in AdS for any dimension and any mass of the complex scalar field. This suggests that the AdS picture captures the underlying dynamics more thoroughly. (Note that the only sphere case where we find the return structure is the one related to AdS$_4$ by a conformal transformation, though the AdS version only includes a subset of $S^3$ modes, see \cite{BCEHLM}.)

Nonlinear wave equations in AdS have been considered in the context of gravitational holography research \cite{Karch:2002sh,Sakai:2004cn,Hartnoll:2008vx,Nishioka:2009zj,Basu:2011ft,Basu:2012gg}. It would be interesting to contemplate whether the weakly nonlinear dynamical return phenomena we have described here have implications from the standpoint of holographic interpretation of AdS dynamics.

We would like to conclude with an even more straightforward connections of our AdS analysis to real-life physics. One can take a nonrelativistic limit of the wave equation (\ref{AdSwave}) in AdS by introducing the would-be nonrelativistic wavefunction $\Psi(t,r,\Omega)$ which is related to the relativistic field $\phi(t,x,\Omega)$ satisfying (\ref{AdSwave}) by
\beq
\phi(t,x,\Omega)=\sqrt{2m}\,e^{-imt}\, \Psi(t, x\sqrt{m},\Omega).
\eeq
One can then check that taking the limit $m\to\infty$ and enforcing the wave equation inside any finite ball in terms of the $r$-coordinate results in the Gross-Pitaevskii equation with a harmonic potential for $\Psi$:
\beq\label{GPHO}
i\frac{\del\Psi}{\del t}=\frac12\left(-\nabla^2+r^2\right)\Psi +|\Psi|^2\Psi.
\eeq
The Gross-Pitaevskii equation describes trapped Bose-Einstein condensates that can be created in a lab using ultracold atomic gases \cite{BEC1,BEC2,BEC3}.
From the above AdS-based construction, it is not surprising that the Gross-Pitaevskii equation (\ref{GPHO}) enjoys a nonrelativistic version of conformal symmetry known as the Schr\"odinger symmetry, as described in \cite{OFN} (a classic treatment of the same symmetry group without the nonlinear term in (\ref{GPHO}) can be found in \cite{Niederer}). Analogies between weakly nonlinear dynamics of the Gross-Pitaevskii equation and AdS systems have been previously pointed out in \cite{BMP}, and the above non-relativistic limit makes the analogy precise for wave equations in AdS. Viewed from this perspective, our present results in AdS generalize the periodic behaviors of the LLL equation described in \cite{ABCE}, since the LLL equation simply represents the maximally rotating sector of the weakly nonlinear dynamics of the Gross-Pitaevskii equation with a harmonic potential.

\section*{Acknowledgments}

This work has been strongly influenced by our collaboration on related subjects and discussions with Piotr Bizo\'n. We furthermore thank Joaquim Gomis for a useful discussion on nonrelativistic limits and kinematic symmetries. Research presented here has been supported in part by the Belgian Federal Science Policy Office through the Interuniversity Attraction Pole P7/37, by FWO-Vlaanderen through projects G020714N and G044016N, and by Vrije Universiteit Brussel (VUB) through the Strategic Research Program ``High-Energy Physics''.
The work of O.E.\ is funded under CUniverse
research promotion project by Chulalongkorn University (grant reference CUAASC).

%%%%%%%%%%%%%%%%%%%%%%%%%%%%%%%%%%%%%%%%%%%%%%%%%%%%

\end{document}